\documentclass[aps, pra, preprint, groupedaddress, amsfonts,
               amsmath, amssymb, nofootinbib]{revtex4-2}
\usepackage{microtype}
\usepackage{graphicx}
\usepackage{epstopdf}
\usepackage[utf8]{inputenc}
\usepackage[T1]{fontenc}
\usepackage[usenames,dvipsnames]{xcolor}
\usepackage{hyperref}
\usepackage{lmodern}
\usepackage{comment}
\usepackage{subcaption}
\usepackage{float}

\usepackage{color}

\begin{document}
\title{Investigations of electron removal processes in slow He$^{2+}$- and He$^{+}$-Ne$_{2}$ collisions and of their implications for the subsequent dimer fragmentation through interatomic Coulombic decay}

\author{Darij Starko}
\email{dmstarko@yorku.ca}
\thanks{Author to whom any correspondence should be addressed.}
\affiliation{Department of Physics and Astronomy, York University, Toronto, Ontario M3J 1P3, Canada}
\author{Tom Kirchner}
\email{tomk@yorku.ca}
\affiliation{Department of Physics and Astronomy, York University, Toronto, Ontario M3J 1P3, Canada}

\date{\today}
\begin{abstract}
We implement an independent-atom and independent-electron model to investigate the collision systems of He$^{2+}$ and He$^{+}$ ion projectiles impinging on a neon dimer target. The dimer is set to be stationary at its equilibrium bond length with the projectile traveling parallel to the dimer axis at a speed corresponding to the collision energy of 10 keV/amu. Two approaches, namely multinomial and determinantal, are used as an analysis of these collisions. Each of the analyses is broken down into two types of models that do not and do include a change in the projectile charge state due to electron capture from the dimer. All calculations are performed using both a frozen atomic target and a dynamic response model using the coupled-channel two-center basis generator method for orbital propagation. All one-electron and two-electron removal processes are calculated, though particular attention is paid to those that result in the Ne$^{+}$-Ne$^{+}$ fragmentation channel due to its association with interatomic Coulombic decay (ICD). For He$^{2+}$ impact, we find that Ne(2$s$) electron removal is strong across all analyses and models, which is in line with previous results that show that ICD contributes to dimer fragmentation through that channel. We also find indications that there is a pure ICD yield when utilizing a He$^{+}$ projectile and applying the model that takes into account the change in projectile charge state.
\end{abstract}

\maketitle
\section{Introduction}
\label{intro}
Rare gas dimers such as Ne$_{2}$, have attracted considerable attention in recent years, in part because the removal of an inner-valence electron is associated with interatomic Coulombic decay (ICD), which results in low-energy electron emission and dimer fragmentation~\cite{Jahnke20}. Rare-gas dimers are compounds formed by weak van der Waals interactions joining a pair of atoms~\cite{najjari2021}. The ICD process is the transfer of energy, typically associated with the formation of a hole, between both partners and the resulting de-excitation removing an electron from the neighbour~\cite{Jahnke20}. Its ubiquity across various research fields coupled with the applied relevance of low-energy electrons owing to their effectiveness in breaking molecular bonds and inflicting damage to biological matter~\cite{Bouda2000} is a significant driver of high activity in the field~\cite{Jahnke20}. 

In most of the experimental works discussed in the review paper~\cite{Jahnke20}, photon fields were used to initiate ICD. Charged-particle collisions present an alternative. Focusing on the prototypical neon dimer target, a positively-charged ion projectile can cause a variety of single and multiple electron removal processes leading to dimer fragmentation. In addition to ICD, which is our main interest here, the fragmentation mechanisms radiative charge transfer (RCT) and direct Coulomb explosion (CE) have been observed~\cite{Iskandar15}.

RCT involves the removal of two valence electrons from one of the two atoms, and then radiative relaxation to a Ne$^{+}(2p^{-1})$+Ne$^{+}(2p^{-1})$ state. In the simplest direct CE process, one valence electron from each atom is removed resulting in the dimer becoming singly charged ions that fragment~\cite{Jahnke20}. Experimentally, ICD, RCT and CE can be distinguished by a careful analysis of the kinetic energy release spectra of the fragmenting ions~\cite{Iskandar2014}, ideally obtained in coincidence with electron energy spectra for an unequivocal identification of ICD~\cite{Jahnke04}. The various electron dynamics that precede these dimer fragmentation mechanisms are studied and presented in this work for 10 keV/amu He$^{+}$ and He$^{2+}$ projectiles.

Experiments carried out by Kim \textit{et al}~\cite{Kim13} for these collision systems at impact energies
between 125 and 165 keV/amu revealed strong low-energy electron emission yields in various fragmentation channels corresponding to ICD. At these higher impact energies, direct target ionization to the continuum is the dominating electron removal process. An energy of 10 keV/amu, by contrast, corresponds to a projectile speed of 0.63 au which is small compared to the average orbital speeds of the inner- and outer-valence electrons of neon. In this \textit{low-energy} regime, electron capture by the projectile dominates and ionization is negligible~\cite{Kirchner_2023}. We believe that low-energy ion-dimer collisions warrant more investigation as 
electron capture, in some systems, is better suited to facilitate ICD~\cite{Iskandar15}. This is due to direct ionization favouring the removal of the most weakly bound electrons, which does not result in ICD. Experimental investigations of the particular low-energy ion-dimer collisions presented in this paper have not yet been conducted and would offer an interesting alternative pathway to ICD production in He$^{2+}$ and He$^{+}$-Ne$_{2}$ collisions. The purpose of this paper is to guide such future experimental efforts.

To model these collisions we make some assumptions, such as the independent atom model (IAM) where each atom in the dimer and its interaction with the projectile can be independently considered in terms of a time-dependent Hamiltonian in the semi-classical approximation~\cite{Kirchner_2021}. For each atom we also use the independent electron model (IEM) framework where electron correlation effects are ignored and instead a single-particle time-dependent Hamiltonian expressed in atomic units with the form 
\begin{equation}
	H(t)=-\dfrac{1}{2}\Delta+\upsilon_{T}\left(r\right)+\upsilon_{P}\left(\boldsymbol{r},t\right)
	\label{eq:hamiltonian}
\end{equation}
is used~\cite{Dyuman20}. $\upsilon_{T}\left(r\right)$ is a spherically symmetric effective target potential, which includes the nuclear Coulomb potential. More specifically, $\upsilon_{T}\left(r\right)$ is obtained from an optimized potential method (OPM) calculation at the level of the exchange-only approximation~\cite{ee93}. The term $\upsilon_{P}\left(\boldsymbol{r},t\right)$ is the bare Coulomb potential of the helium nucleus in the case of He$^{2+}$ impact and the sum of that Coulomb potential and a screening potential of the form
\begin{equation}
	\upsilon_{scr}\left(r_{p}\right)=\int d^{3}r_{p}'\dfrac{\varphi_{1s}^{2}\left(r_{p}'\right)}{\left|\vec{r_{p}}-\vec{r_{p}'}\right|}
	\label{eq:scr_pot}
\end{equation}
in the case of He$^{+}$. In equation (\ref{eq:scr_pot}) $r_{p}=\left|\vec{r}-\vec{R}\left(t\right)\right|$, where $\vec{R}\left(t\right)$ is the classical projectile trajectory and $\varphi_{1s}$ is the (numerical) ground-state Hartree-Fock (HF) orbital of neutral helium. This choice ensures that an initial target electron captured into the ground-state of the projectile has the HF binding energy of neutral helium ($\varepsilon_{1s}=-0.918$ au). An alternative choice, used e.g. in~\cite{Kirchner_2001} would be the hydrogen-like $1s$ orbital of He$^{+}$, which results in the less realistic binding energy $\varepsilon_{1s}'=-0.821$ au of a captured electron. Both choices, of course, result in a $\upsilon_{P}$ that goes as $-2r^{-1}$ at close distances and decays like $-1/r$ asymptotically, and do not lead to significantly different results for the removal processes of interest in this work. 

The ion-atom problem is formulated in the two-center basis generator method (TC-BGM), which is a method of generating basis states with respect to the two centers of both the target and projectile to solve the time-dependent Schrödinger equation~\cite{tcbgm}. The basis used in the present work consists of the $2s$ to $4f$ target orbitals as well as the $1s$ to $7i$ projectile orbitals and a set of pseudo-states~\cite{Kirchner_2021}. The transition amplitudes for electrons changing states are obtained from solving the $N$ single-electron time-dependent Schrödinger equations for the single-particle Hamiltonian (\ref{eq:hamiltonian}). 

Atomic orbitals with real instead of complex spherical harmonics are used in practice such that the orbitals are categorized as even (gerade) and odd (ungerade)~\cite{Dyuman20}. This has the benefit that one avoids the mixing of gerade and ungerade states which simplifies the subsequent calculations.

We build upon previous work~\cite{Kirchner_2021}, and introduce further complexity and alternate approaches for an in-depth study of the ion-dimer collisions. Our assumptions include a stationary neon dimer during the collision at a fixed bond length of $R_e=5.86$ a.u. The projectile travels at constant speed $v=|\bf v|$ with an energy of $E_P = 10$ keV/amu in a straight line and travels parallel to the dimer axis, maintaining the same impact parameters between it and the two target atoms during the collision. In the parallel orientation, we treat projectile collisions with dimer atoms in a sequential manner, which is a key feature we exploit in the models we develop. 

In~\cite{Kirchner_2021}, a multinomial, combinatoric model of electron removal, and its parts 
ionization and capture, were set up and applied to He$^{2+}$-Ne$_{2}$ collisions in an impact energy
range from 2.81 to 200 keV/amu. Three perpendicular orientations of the dimer with respect to the projectile beam axis were considered in order to compute orientation-averaged cross sections for various removal processes. Data were compared to the measurements of Kim \textit{et al}~\cite{Kim13} at 150 keV/amu and found to be consistent with them. We here expand on this model to incorporate the effects of electron capture by the projectile and also to account for the antisymmetry of the many-electron initial and final states. Given the nature of our capture model (discussed in section~\ref{sec:model}) we restrict the present analysis to 10 keV/amu impact energy and the parallel ion-dimer orientation. We investigate the possibility of maximizing the ICD yield\footnote{The ICD yield is the ratio of the $2s^{-1}$ cross-section over the sum of all possible electron removal channels that potentially result in Ne$^{+}$+Ne$^{+}$ fragmentation.} through the use of He$^{2+}$ and He$^{+}$ ions as projectiles. The models that examine this are summarized in section~\ref{sec:model}. Results are discussed in section~\ref{sec:Results} and conclusions are offered in section~\ref{sec:conclusions}.

\section{Model}
\label{sec:model}
In a deviation from the analysis of~\cite{Kirchner_2021}, we develop a model that addresses the change in projectile charge by electron capture, which we call the capture model. In this model, we make the assumption that the projectile interacts with each atom independently in a sequential manner where electron capture from the first atom changes the collision dynamics with the second atom. For this reason only the parallel orientation will be studied in which this sequential scenario is more realistic than in other geometries. We note that orientation-dependent information is accessible experimentally via measurement of the momenta of the dimer fragments (see, e.g.,~\cite{Siddiki_2023,Kim2014,Iskandar2014} and references cited therein), i.e., a future experimental study that directly compares with our results for one particular orientation seems feasible in principle.

To address the effects of electron capture by the projectile, comparisons between models that do and do not take this into account are analysed, which we call capture and fixed-charge respectively. Two analyses are carried out: The first is a multinomial approach which combines single-electron probabilities for capture or no removal and is derived in a combinatorial way that does not explicitly account for the Pauli exclusion principle. The second approach is through a determinantal analysis, in which determinants of single-particle density matrices are computed and which does account for the Pauli exclusion principle. The electron removal processes from the dimer are broken down into three types: one electron removal, one electron removal from each atom in the dimer (two-site), and two electron removal from a single atom (one-site).

The mentioned analyses are further broken down by considering them with a frozen target potential (no-response) model and a dynamic response model following the details laid out in~\cite{tom00}. The idea behind the response model is to adjust the atomic potential in the spirit of time-dependent mean-field theory to reflect the fractional ionic character a target atom acquires when electrons are removed from it. The ionic character is regulated by the time-dependent net electron removal probability. The response effects remain marginal so long as zero- and one-fold electron removal dominate, but they may become significant once higher-order removal processes contribute substantially.

The dynamical response approach for each analysis and model is conducted as the no-response model, except that the final-state analyses are carried out at eleven, instead of just one, final distances between the projectile and the target, with results averaged. Due to the dynamically changing potential, projecting the solutions onto inital target eigenstates results in fluctuating transition probabilities~\cite{tom00}. This then requires calculating these probabilities by sampling a set of distances (ranging from $z=30$ a.u. to $50$ a.u. in the present work) and then averaging~\cite{Schenk15}.  

\subsection{Multinomial Analysis}
\label{ssec:MA}
In the multinomial analysis, a straightforward approach is chosen whereby the products of single-particle probabilities in ion-atom collisions are combined to achieve a particular end state (channel). To understand this approach, the simplest removal of a single $2s$ electron from one neon atom in the dimer is analysed, with electron capture by the projectile not changing the interactions between it and the target:
\begin{equation}
	P_{2s^{-1}}^{\rm rem}(b) =
	4 p_{2s}^{\rm rem}(b) [1-p_{2s}^{\rm rem}(b)]^3 [1-p_{2p_0}^{\rm rem}(b)]^4 
	[1-p_{2p_{1g}}^{\rm rem}(b)]^4 [1-p_{2p_{1u}}^{\rm rem}(b)]^4.
	\label{eq:picd}
\end{equation}
The variable $p_{2s}^{\rm rem}(b)$ corresponds to a removal probability of a $2s$ electron, where $2s$, $2p_{0}$, $2p_{1g/u}$ are the electron states under consideration and $g/u$ stands for gerade/ungerade. We assume that direct ionization is so weak a process at 10 keV/amu that it can be ignored, i.e. we identify removal with capture. We also assume the $K$-shell electrons to be passive. The factor of four in equation (\ref{eq:picd}) corresponds to the four possible $2s$ electrons of the dimer that can be captured. Using this information, $(1-p_{2s}^{\rm rem}(b))$ is the probability that an electron remains bound to the target. Given that target electron excitation is a weak process, we can identify it with the elastic probability. The product of probabilities can be easily understood in terms of the projectile's interactions with each dimer atom in its trajectory. 

The multinomial analysis is then modified to account for the projectile's change in charge state due to electron capture, which constitutes the capture model. This model recalculates the probability terms in the case of a He$^{+}$ projectile. The same method as before is used, though accounting for the captured electron by the projectile, the probability terms corresponding to that interaction are used instead. For notational convenience, we use $*$ to denote He$^{+}$ interaction probability terms (e.g. $p_{2p_{0}}^{*}$, $1-p_{2p_{0}}^{*}$). The resulting $2s$ electron removal probability in the capture model is
\begin{equation}
\begin{aligned}
	P_{2s^{-1}} ={} &
	2 p_{2s} [1-p_{2s}] [1-p_{2p_0}]^2 
	[1-p_{2p_{1g}}]^2 [1-p_{2p_{1u}}]^2 
	[1-p_{2s}^{*}]^2 [1-p_{2p_0}^{*}]^2 
	[1-p_{2p_{1g}}^{*}]^2 [1-p_{2p_{1u}}^{*}]^2 \\
	&+ 2 p_{2s} [1-p_{2s}]^3 [1-p_{2p_0}]^4 
	[1-p_{2p_{1g}}]^4 [1-p_{2p_{1u}}]^4 .
	\label{eq:pcicd}
\end{aligned}
\end{equation} 
The superscript $\rm rem$ and the argument $(b)$ are suppressed out of convenience. The first term on the right-hand side of equation (\ref{eq:pcicd}) corresponds to the removal probability of a single electron from the first atom, while the second term corresponds to the electron removal probability from the second atom. For the first term we note that when an electron is removed from the first atom, the change in projectile charge results in different probability terms for the elastic interaction with the second atom. In the limit that $p_{2s}=p_{2s}^{*}$, we recover the fixed-charge model equation (\ref{eq:picd}). If the projectile has captured two electrons from the first atom it encounters it becomes neutral and its interaction with the second atom cannot result in additional electron removal, i.e. the elastic probabilities become equal to one. Similarly, an initial He$^{+}$ projectile will only be able to capture a single electron in the capture model. The $2s$ electron removal probability in the capture model is then:

\begin{equation}
	\begin{aligned}
		P_{2s^{-1}} ={} &
		\left(1+[1-p_{2s}]^2 [1-p_{2p_0}]^2 
		[1-p_{2p_{1g}}]^2 [1-p_{2p_{1u}}]^2\right)  \\
		& 2 p_{2s} [1-p_{2s}] [1-p_{2p_0}]^2 
		[1-p_{2p_{1g}}]^2 [1-p_{2p_{1u}}]^2 .
		\label{eq:hepcicd}
	\end{aligned}
\end{equation}

We see in equation (\ref{eq:hepcicd}) that the electron removal probability by the neutral He atom is set equal to zero. These features of the capture model are consistent with our assumption that direct ionization is negligibly small for all projectile charge states and with neglecting the formation of (metastable) negative helium ions in addition~\cite{Reinhed_2009}.

\subsection{Determinantal Analysis}
\label{ssec:DA}
In this analysis, we assume that the transition amplitudes are described by inner products of Slater determinants, that by design account for the Pauli exclusion principle. The transition from initial-to-final-state electron configurations results in a probability that is equal to the determinant of a single-particle density matrix:
\begin{equation*}
	P_{f_{1}\ldots f_{N}} = |\left\langle f_{1}\ldots f_{N}|i_{1}\ldots i_{N},t_{f}\right\rangle|^2
\end{equation*}
\begin{equation}
	\begin{aligned}
	& =
	\begin{vmatrix}
		\left\langle f_{1}|i_{1}\right\rangle & \cdots & \left\langle f_{1}|i_{N}\right\rangle \\
		\vdots        & \ddots & \vdots \\
		\left\langle f_{N}|i_{1}\right\rangle & \cdots & \left\langle f_{N}|i_{N}\right\rangle
	\end{vmatrix}
	\times
		\begin{vmatrix}
		\left\langle i_{1}|f_{1}\right\rangle & \cdots & \left\langle i_{1}|f_{N}\right\rangle \\
		\vdots        & \ddots & \vdots \\
		\left\langle i_{N}|f_{1}\right\rangle & \cdots & \left\langle i_{N}|f_{N}\right\rangle
	\end{vmatrix} \\
	& = 
	\begin{vmatrix}
		\gamma_{11} & \cdots & \gamma_{1N} \\
		\vdots        & \ddots & \vdots \\
		\gamma_{N1} & \cdots & \gamma_{NN}
	\end{vmatrix} \equiv det(\gamma).
	\end{aligned}
	\label{eq:exclusive}
\end{equation}
Here $\left|i_{j},t_{f}\right\rangle$ is the time-propagated orbital corresponding to the initial state $\left|i_{j}\right\rangle$, $\left|f_{j}\right\rangle$ is a final state, $\gamma_{jk}$ are the density matrix elements computed from the Slater determinants, and $N$ is the number of electrons ($N$=8 in the present work). The density matrix elements can be written as sums of products of single-particle transition amplitudes from a range of initial to final states:
\begin{equation}
	\gamma_{jk}(t_{f}) = \left\langle f_{j}|\gamma\left(t_{f}\right)|f_{k}\right\rangle =\sum_{i=1}^{N}\left\langle f_{j}|i,t_{f}\right\rangle \left\langle i,t_{f}|f_{k}\right\rangle =\sum_{i}^{N}c_{k}^{{i}^{*}}\left(t_{f}\right)c_{j}^{i}\left(t_{f}\right).
\end{equation}

In a situation where we explicitly dictate the final states of all the electrons in the system, as seen in equation (\ref{eq:exclusive}), we call this the exclusive probability. This probability corresponds to a complete measurement. However, this is in general not realistically possible and typically we do not have all this information. The probability corresponding to transitions of some electrons while not firmly choosing all final states, is called the inclusive probability~\cite{Ludde_1985}. This probability is a sum of exclusive probabilities of some electron states and can be shown to equal:
\begin{equation}
	P_{f_{1}\ldots f_{q}}=\sum_{f_{q+1}<\ldots<f_{N}}P_{f_{1}\ldots f_{N}}.
	\label{eq:inclusive}
\end{equation}
Here $q$ out of $N$ electrons are in the final state $\left|f_{1}\ldots f_{q}\right\rangle$ while nothing is known about the other $N-q$ electrons. Consistent with equation (\ref{eq:exclusive}), $P_{f_{1}\ldots f_{q}}$ is the determinant of one $q\times q$ matrix corresponding to the sub-configuration $\left|f_{1}\ldots f_{q}\right\rangle$. When electrons are removed, it is also important to understand what happens with the emptied spaces they occupied, the holes. Thus, using the structure of (\ref{eq:inclusive}), we formulate an inclusive particle-hole probability as follows:
\begin{equation}
	\begin{aligned}
		\begin{array}{c}
			P_{f_{1}\ldots f_{q}}^{\bar{f_{1}}\ldots\bar{f_{k}}}\equiv\sum_{f_{q+1}<\ldots<f_{N}}^{\bar{f_{1}}\ldots \bar{f_{k}}}P_{f_{1}\ldots f_{N}}\\
			=P_{f_{1}\ldots f_{q}}-{\displaystyle \sum_{l=1}^{k}}P_{f_{1}\ldots f_{q}\bar{f_{l}}}+{\displaystyle \sum_{l_{1}<l_{2}}^{k}}P_{f_{1}\ldots f_{q}\bar{f_{l_{1}}}\bar{f_{l_{2}}}}-\cdots,
		\end{array}
	\end{aligned}
	\label{eq:holes}
\end{equation}
where $\bar{f_{i}}$ are the hole states and the particle-hole probability is an alternating sum of positive and negative terms~\cite{Ludde_1985,AToepfer_1985}. This formalism can be used to find the probabilities of the electron removal channels of interest in this work, as will be detailed further below.

The inclusive model can be extended to include the projectile ion in the analysis. For He$^{2+}$ impact, we consider two models that take into account the capture of an electron by the projectile to form an He$^{+}$ ion and its interaction with the second atom in the dimer; models I and II.

Model I incorporates the above inclusive probability formulation, yet using transition amplitudes computed from interactions between He$^{+}$ and an Ne atom in addition to those for He$^{2+}-$Ne collisions. The channel probabilities are then changed in an analogous way as in the multinomial capture model, while following the determinantal fixed-charge model approach. 

Model II has fundamentally the same structure as model I when computing final probabilities, however the density matrix elements include initial states from the projectile ion. We only consider the $1s$ state, due to that being the most relevant state an electron will occupy in the ion after capture. We can account for this by writing
\begin{equation}
	\gamma_{kj}=\sum_{i=1}^{N+1}\left\langle f_{k}|i,t_{f}\right\rangle \left\langle i,t_{f}|f_{j}\right\rangle.
\end{equation}

The inclusion of the transition from the projectile to the second target atom is indicated by the +1 above the sum.

We can see how this analysis and its models are applied when again inspecting, as we did in the multinomial case, the $2s$ electron removal probability as an example.

First we start with the fixed-charge model. As in the multinomial analysis, we ignore the change in the projectile charge after capturing a removed electron. We recognize that the $2s$ electron removal is a sum of particle-hole probabilities, and start with the removal of a $2s$ electron from one neon atom:
\begin{equation}
P_{2s^{-1}}^{Ne}=P_{2s\downarrow,2p_{0}^{2},2p_{1g}^{2},2p_{1u}^{2}}^{2\bar{s}\uparrow}+P_{2s\uparrow,2p_{0}^{2},2p_{1g}^{2},2p_{1u}^{2}}^{2\bar{s}\downarrow}.
\label{eq:det_i_ne}
\end{equation}
The arrows indicate the spin of the electron and hole. There are two terms to account for the fact that either a spin-up or a spin-down electron can be removed. We then apply (\ref{eq:holes}) in (\ref{eq:det_i_ne}) and obtain for the first term
\begin{equation}
P_{2s\downarrow,2p_{0}^{2},2p_{1g}^{2},2p_{1u}^{2}}^{2\bar{s}\uparrow}=P_{2s\downarrow,2p_{0}^{2},2p_{1g}^{2},2p_{1u}^{2}}-P_{2s^{2},2p_{0}^{2},2p_{1g}^{2},2p_{1u}^{2}}.
\label{eq:det_i_1st_term}
\end{equation}
The first term on the right-hand side of equation (\ref{eq:det_i_1st_term}) corresponds to having a spin-down $2s$ electron along with a filled $2p$ subshell on the neon atom, while the final state of one electron is not determined. The second term is the elastic probability, which is the probability for all electrons to remain in place. We neglect target excitation and direct ionization, which is justified in the considered scenario so that the difference of both terms corresponds to the probability to find a hole in the target $2s\uparrow$ state and the missing electron captured by the projectile. 

Inserting (\ref{eq:det_i_1st_term}) into (\ref{eq:det_i_ne}), we obtain for $2s$ electron removal:
\begin{equation}
P_{2s^{-1}}^{Ne}=P_{2s\downarrow,2p_{0}^{2},2p_{1g}^{2},2p_{1u}^{2}}-P_{2s^{2},2p_{0}^{2},2p_{1g}^{2},2p_{1u}^{2}}+P_{2s\uparrow,2p_{0}^{2},2p_{1g}^{2},2p_{1u}^{2}}-P_{2s^{2},2p_{0}^{2},2p_{1g}^{2},2p_{1u}^{2}}.
\label{eq:full_det_i_ne}
\end{equation}
We know that the Hamiltonian is spin independent, i.e., the transition amplitudes are identical for both spin-up and spin-down electrons. Accordingly, the inclusive probabilities with a spin-down and spin-up electron present are identical as well, resulting in the final expression
\begin{equation}
P_{2s^{-1}}^{Ne}=2\left(P_{2s,2p_{0}^{2},2p_{1g}^{2},2p_{1u}^{2}}-P_{\rm elastic}\right)
\label{eq:final_det_i_ne}
\end{equation}
with 
\begin{equation}
	P_{\rm elastic}=P_{2s^{2},2p_{0}^{2},2p_{1g}^{2},2p_{1u}^{2}}.
	\label{eq:elastic}
\end{equation}
In equation (\ref{eq:final_det_i_ne}), we dropped the arrow beside the subscript $2s$ due to the identical spin terms.
Within the independent-atom model we can construct all the one- and two-site electron removals from the dimer using the inclusive probabilities as follows:
\begin{equation}
\begin{array}{c}
	P_{s^{-1}}^{\rm dimer}=2P_{s^{-1}}P_{\rm elastic}\\
	P_{p^{-1}}^{\rm dimer}=2P_{p^{-1}}P_{\rm elastic}\\
	P_{s^{-1},s^{-1}}^{\rm dimer}=P_{s^{-1}}P_{s^{-1}}\\
	P_{p^{-1},p^{-1}}^{\rm dimer}=P_{p^{-1}}P_{p^{-1}}\\
	P_{s^{-1},p^{-1}}^{\rm dimer}=2P_{s^{-1}}P_{p^{-1}}\\
	P_{s^{-2}}^{\rm dimer}=2P_{s^{-2}}P_{\rm elastic}\\
	P_{p^{-2}}^{\rm dimer}=2P_{p^{-2}}P_{\rm elastic}\\
	P_{s^{-1}p^{-1}}^{\rm dimer}=2P_{s^{-1}p^{-1}}P_{\rm elastic}.
\end{array}
\label{eq:int_p}
\end{equation}

The subscripts in (\ref{eq:int_p}) have the 2 dropped for convenience and on the right-hand sides the superscript Ne has been omitted as well. The terms $s^{-1},s^{-1}$ and $s^{-2}$ correspond to two-site and one-site $2s$ electron removal, as the comma here implies two-site. This follows for the rest of the electron removal channels listed in (\ref{eq:int_p}).

We now apply what we have done to the capture model I. The channel probabilities are changed in an identical way as in the multinomial capture model, and we obtain
\begin{equation}
\begin{array}{c}
	P_{s^{-1}}^{\rm dimer}=P_{s^{-1}}P_{\rm elastic}^{*}+P_{\rm elastic}P_{s^{-1}}\\
	P_{p^{-1}}^{\rm dimer}=P_{p^{-1}}P_{\rm elastic}^{*}+P_{\rm elastic}P_{p^{-1}}\\
	P_{s^{-1},s^{-1}}^{\rm dimer}=P_{s^{-1}}P_{s^{-1}}^{*}\\
	P_{p^{-1},p^{-1}}^{\rm dimer}=P_{p^{-1}}P_{p^{-1}}^{*}\\
	P_{s^{-1},p^{-1}}^{\rm dimer}=P_{s^{-1}}P_{p^{-1}}^{*}+P_{p^{-1}}P_{s^{-1}}^{*}\\
	P_{s^{-2}}^{\rm dimer}=P_{s^{-2}}+P_{\rm elastic}P_{s^{-2}}\\
	P_{p^{-2}}^{\rm dimer}=P_{p^{-2}}+P_{\rm elastic}P_{p^{-2}}\\
	P_{s^{-1}p^{-1}}^{\rm dimer}=P_{s^{-1}p^{-1}}+P_{\rm elastic}P_{s^{-1}p^{-1}}.
	\label{eq:cap_m0_p}
\end{array}
\end{equation}

The last three probabilities in (\ref{eq:cap_m0_p}) assume that interactions between a neutral He atom and Ne will be negligible.

As mentioned earlier, model II has fundamentally the same structure as model I when computing final probabilities, however the density matrix elements include the $1s$ projectile state. For example, the density matrix element $\gamma_{2s\uparrow2p_{0}\uparrow}$ would be represented in full by
\begin{equation} 
\gamma_{2s\uparrow2p_{0}\uparrow}=c_{2s}^{1s(P)}c_{2p_{0}}^{1s(P)*}+c_{2s}^{2s(T)}c_{2p_{0}}^{2s(T)*}+c_{2s}^{2p_{0}(T)}c_{2p_{0}}^{2p_{0}(T)*}+c_{2s}^{2p_{1g}(T)}c_{2p_{0}}^{2p_{1g}(T)*},
\end{equation}
where $(P)$ and $(T)$ represent projectile and target initial states respectively. This formulation is correct if we fix the projectile $1s$ state to be in the spin up orientation. For a spin down density matrix element, the first term on the right-hand side would vanish, i.e., $\gamma_{2s\uparrow2p_{0}\uparrow}$ is different from $\gamma_{2s\downarrow2p_{0}\downarrow}$ due to this asymmetry.

The impact of this inclusion is seen in a simple example where all $2p$ states are ignored, and we consider the removal of one $2s$ electron. Assuming that the electron in the He$^{+}$ ion is in a spin up state, the removal probability from a single atom is
\begin{equation}
P_{2s}^{-1}=(P_{2s\uparrow}-P_{2s^{2}})+(P_{2s\downarrow}-P_{2s^{2}}),
\label{eq:det_full_p}
\end{equation}
where $P_{2s\uparrow}$ and $P_{2s\downarrow}$ are the inclusive probabilities to remove a spin-down and spin-up electron from the target respectively. As we have fixed the electron spin in the He$^{+}$ ion to be in the up state, this electron cannot transition to a spin down state on the target and vice-versa. We thus see an asymmetry in spin-up and down removals, and formulate two different probability terms using the density matrix element formulation as follows
\begin{equation}
P_{2s\uparrow}=c_{2s\uparrow}^{1s\uparrow(P)}c_{2s\uparrow}^{1s\uparrow(P)*}+c_{2s\uparrow}^{2s\uparrow(T)}c_{2s\uparrow}^{2s\uparrow(T)*}+c_{2s\uparrow}^{2s\downarrow(T)}c_{2s\uparrow}^{2s\downarrow(T)*},
\label{eq:det_m1_p}
\end{equation}
where the last term is 0 given that spin flips are not permitted. Using this same logic we obtain for the spin-down term,
\begin{equation}
P_{2s\downarrow}=c_{2s\downarrow}^{2s\downarrow(T)}c_{2s\downarrow}^{2s\downarrow(T)*}.
\label{eq:spin_down}
\end{equation}
In the case of the probability $P_{2s^{2}}$ we obtain
\begin{equation}
P_{2s^{2}}=
\left|\begin{array}{cc}
	P_{2s\uparrow}  & 0\\
	0 & P_{2s\downarrow}
\end{array}\right|.
\label{eq:slater}
\end{equation}
Taking the determinant and substituting equations (\ref{eq:det_m1_p}) and (\ref{eq:spin_down}) into (\ref{eq:slater}) we obtain
\begin{equation}
P_{2s^{2}}=\left(c_{2s\uparrow}^{1s\uparrow(P)}c_{2s\uparrow}^{1s\uparrow(P)*}+c_{2s\uparrow}^{2s\uparrow(T)}c_{2s\uparrow}^{2s\uparrow(T)*}\right)\left(c_{2s\downarrow}^{2s\downarrow(T)}c_{2s\downarrow}^{2s\downarrow(T)*}\right).
\end{equation}
Inserting these three probability terms back into (\ref{eq:det_full_p}), we obtain  
\begin{equation}
\begin{aligned}
P_{2s^{-1}}={} & \left(c_{2s\uparrow}^{1s\uparrow(P)}c_{2s\uparrow}^{1s\uparrow(P)*}+c_{2s\uparrow}^{2s\uparrow(T)}c_{2s\uparrow}^{2s\uparrow(T)*}-\left(c_{2s\uparrow}^{1s\uparrow(P)}c_{2s\uparrow}^{1s\uparrow(P)*}+c_{2s\uparrow}^{2s\uparrow(T)}c_{2s\uparrow}^{2s\uparrow(T)*}\right)\left(c_{2s\downarrow}^{2s\downarrow(T)}c_{2s\downarrow}^{2s\downarrow(T)*}\right)\right) \\
&+\left(c_{2s\downarrow}^{2s\downarrow(T)}c_{2s\downarrow}^{2s\downarrow(T)*}-\left(c_{2s\uparrow}^{1s\uparrow(P)}c_{2s\uparrow}^{1s\uparrow(P)*}+c_{2s\uparrow}^{2s\uparrow(T)}c_{2s\uparrow}^{2s\uparrow(T)*}\right)\left(c_{2s\downarrow}^{2s\downarrow(T)}c_{2s\downarrow}^{2s\downarrow(T)*}\right)\right) \\
= {} & 2c_{2s\uparrow}^{2s\uparrow(T)}c_{2s\uparrow}^{2s\uparrow(T)*}\left(1-c_{2s\uparrow}^{2s\uparrow(T)}c_{2s\uparrow}^{2s\uparrow(T)*}\right)+c_{2s\uparrow}^{1s\uparrow(P)}c_{2s\uparrow}^{1s\uparrow(P)*}\left(1-2c_{2s\uparrow}^{2s\uparrow(T)}c_{2s\uparrow}^{2s\uparrow(T)*}\right).
\end{aligned}
\label{eq:final_m1_p}
\end{equation}
We can see the asymmetric impact when including the initial transition from the projectile ion. For completeness, when we consider the case of an electron in the spin-down orientation in the He$^{+}$ ion, we obtain the same end result and if we neglect transitions from the initial projectile electron ($c_{2s\uparrow}^{1s\uparrow(P)}=0$) we recover model I. These expressions, in their generalized forms (including the $2p$ electrons) are then inserted into the electron removal probabilities (\ref{eq:cap_m0_p}) discussed above.

For a He$^{+}$ projectile, only single electron removal processes are allowed in the capture model for the  reasons stated in section~\ref{ssec:MA}. We obtain

\begin{equation}
	\begin{array}{c}
		P_{s^{-1}}^{\rm dimer}=P_{s^{-1}}+P_{\rm elastic}P_{s^{-1}}\\
		P_{p^{-1}}^{\rm dimer}=P_{p^{-1}}+P_{\rm elastic}P_{p^{-1}},\\
		\label{eq:hecap_m0_p}
	\end{array}
\end{equation}

where the first terms on the right-hand side correspond to capture from the first atom encountered, turning the projectile into a neutral atom, while the second terms describe capture form the second atom.

\section{Results and discussion}
\label{sec:Results}
We first consider the He$^{2+}$ projectile, then He$^{+}$, and we focus on the $2s^{-1}$ channel which is associated with ICD. Going forward, in all figures references to electron channels will not include the principal quantum number $n=2$ in front (i.e. we replace $2s$ with $s$ etc.); res will refer to the response model and hep to the He$^{+}$ projectile being used in the analysis. To simplify references with respect to analyses, we refer to the multinomial analysis as MA, and the determinal analysis as DA. Models for an analysis also use acronyms (e.g. fixed-charge model is FCM; capture model I is CMI). Identifying a particular analysis and model uses a XX.YYY format, where XX refers to the analysis and YYY the model (e.g. MA.FCM stands for the fixed-charge model in the multinomial analysis).

\subsection{He$^{2+}$ Projectiles}

\begin{figure}[h!]
	\centering
	\includegraphics[width=0.55\linewidth]{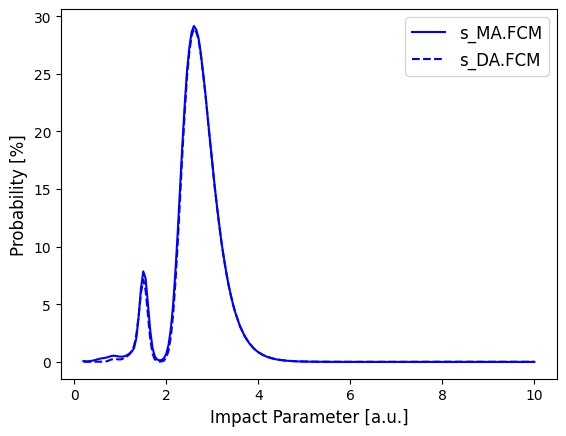}
	\caption{He$^{2+}$-Ne$_{2}$ ion-dimer collisions in parallel orientation at 10 keV/amu; Single $2s$ electron removal probability comparing MA.FCM (solid line) vs DA.FCM (dashed line).}
	\label{fig1}
\end{figure}

To build on the previous work~\cite{Kirchner_2021}, we start by looking at the difference between MA.FCM and DA.FCM, focusing on single $2s$ electron removal from the dimer. 
As seen in figure~\ref{fig1}, the two analyses give almost indistinguishable results, demonstrating that the Pauli principle does not play a significant role for this single vacancy production channel.

\begin{figure}[h!]
	\centering
	\includegraphics[width=0.55\linewidth]{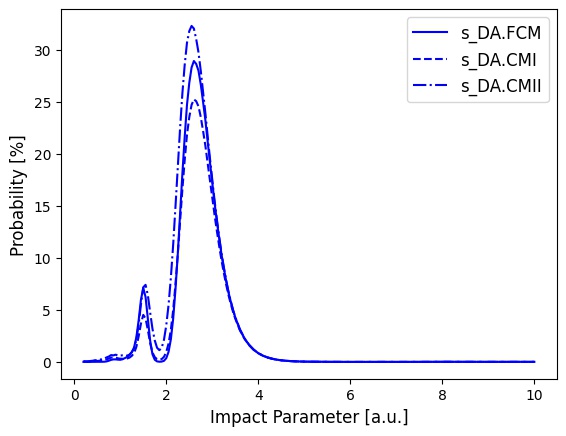}
	\caption{He$^{2+}$-Ne$_{2}$ ion-dimer collisions in parallel orientation at 10 keV/amu; Single $2s$ electron removal probability comparisons of DA.FCM (solid line) and DA.CMI (dashed line) and DA.CMII (dashed-dotted line).}
	\label{fig2}
\end{figure}

This allows us to further explore the DA and compare the two capture models with the fixed-charge model. This is done in figure~\ref{fig2} for the same channel. We see that the capture models are both similar to the fixed-charge model. There is fairly close agreement even when we account for changes in projectile charge due to collision order. Inspecting the first equation in (\ref{eq:cap_m0_p}) we can see that the only term affected by the capture models is $P_{elastic}^{*}$. In model II, transitions of the projectile electron to a target state are included in addition to the strictly elastic term so that the total probability is somewhat larger.

The differences between these models and analyses are also of interest when examining other electron removal channels. These include the removal of a single electron from each atom in the dimer, which we call two-site electron removal. As observed in figure~\ref{midi,mcdc-2site}, we see greater differences in the $2s^{-1}$,$2s^{-1}$ channel for the capture model that are not observed for the single $2s$ electron removal channel. This corresponds to the effects of electron capture by the projectile, making it less likely for secondary electron capture. However, this is not the case for the $2s^{-1}$,$2p^{-1}$ and $2p^{-1}$,$2p^{-1}$ channels due to easier removal of $2p$ electrons with a He$^{+}$ projectile. Thus the $2s^{-1}$,$2s^{-1}$ channel is interesting, as the difference between the fixed-charge and capture models is substantial, making it a good way to test the capture models when experimental data become available for comparison.

\begin{figure}[h!]
	\centering
	\includegraphics[width=0.55\linewidth]{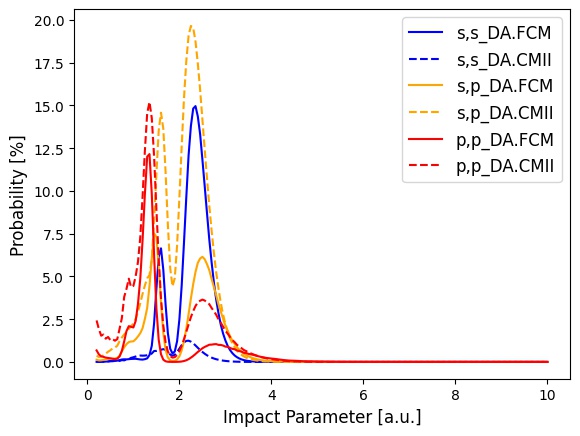}
	\caption{He$^{2+}$-Ne$_{2}$ ion-dimer collisions in parallel orientation at 10 keV/amu; Two-site two electron electron removal probability comparisons of DA.FCM (solid lines) and DA.CMII (dashed lines).}
	\label{midi,mcdc-2site}
\end{figure}

\begin{figure}[h!]
	\centering
	\includegraphics[width=0.55\linewidth]{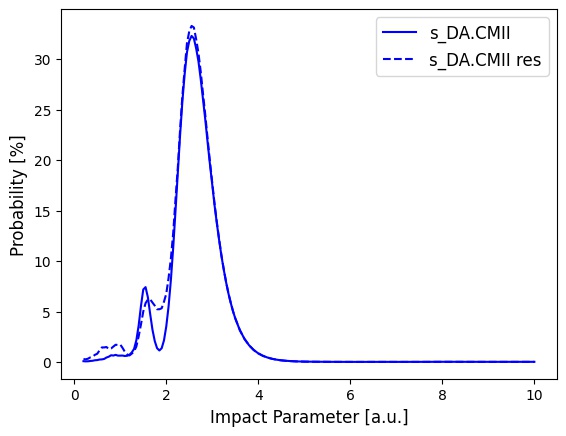}
	\caption{He$^{2+}$-Ne$_{2}$ ion-dimer collisions in parallel orientation at 10 keV/amu; Single $2s$ electron removal probability comparison of DA.CMII between no-response (solid line) and response (res) (dashed line) models.}
	\label{fig3}
\end{figure}

It is also important to consider the consequences of including dynamical response. As seen in figure~\ref{fig3}, the differences between the results obtained with and without the response model, for the $2s^{-1}$ channel with the DA.CMII are small. This is expected given that the response model is designed in a way to mainly affect higher charge states~\cite{tom00}. Indeed, we are seeing larger effects in some of the two-electron removal channels (see table~\ref{table1}) and expect response to gain in importance for multiplicities $q>2$.

Ultimately, we are interested in the total cross-section for a collision as it is the most relevant information. The cross-sections for the removal processes of interest are presented in table~\ref{table1}. They are obtained by the usual integration over the impact parameter range.
\begin{table}
	\begin{center}
		\begin{tabular}{||c | c | c | c | c | c || c ||} 
			\hline
			Models & 2s$^{-1}$ & 2s$^{-1}$,2s$^{-1}$ & 2s$^{-1}$,2p$^{-1}$ & 2p$^{-1}$,2p$^{-1}$ & 2p$^{-2}$ &ICD yield \\ [0.5ex] 
			\hline\hline
			MA.FCM
			& 1.28 & 0.44 & 0.26 & 0.17 & 0.12 & 0.57 \\ 
			\hline
			DA.FCM
			& 1.26 & 0.41 & 0.28 & 0.16 & 0.11 & 0.57 \\
			\hline
			DA.CMII
			& 1.47 & 0.04 & 0.73 & 0.33 & 0.48 & 0.48 \\ 
			\hline
			DA.CMII res
			& 1.55 & 0.05 & 0.72 & 0.37 & 0.56 & 0.48 \\ 
			\hline
		\end{tabular}
		\caption{Total cross-sections in units of\textup{~\AA}$^{2}$ for various models and electron removal processes for the He$^{2+}$-Ne$_{2}$ ion-dimer system. ICD yield is the ratio of the $2s^{-1}$ cross-section over the sum of all presented channels that potentially result in Ne$^{+}$+Ne$^{+}$ fragmentation.}
		\label{table1}
	\end{center}
\end{table}

The last column labeled ICD yield is the ratio of the $2s^{-1}$ cross-section over all processes cross-sections that result in Ne$^{+}$+Ne$^{+}$ fragmentation. As can be seen, the predicted ICD yield is strong, albeit somewhat reduced by the capture model. The two-site two electron channels all lead to CE, and are included in the calculation for the ICD yield for that reason. The channel for $2p^{-2}$ is also included in the table due to the chance of the system Ne$^{2+}(2p^{-2})$+Ne decaying into Ne$^{+}$+Ne$^{+}$ via RCT. For simplicity, we assume that RCT happens with certainty in the $2p^{-2}$ channel but we note that this is debatable (see, e.g., Ref.~\cite{Kirchner_2021}). The channels $2s^{-2}$ and $2s^{-1}2p^{-1}$ are not included as they facilitate Ne$^{2+}$+Ne$^{+}$ fragmentation through ICD~\cite{Kim13}. These two channels, although omitted in this work, are interesting as they are the only allowed three-electron processes in the capture model, meaning the model would predict a pure ICD yield in the Ne$^{2+}$+Ne$^{+}$ fragmentation channel. 

\subsection{He$^{+}$ Projectiles}

The unique feature of considering the He$^{+}$ projectile is that in the capture model we are restricted to the removal of only a single electron from the dimer. This is due to the projectile turning neutral and no longer being able to effectively capture any more electrons.

\begin{figure}[h!]
	\centering
	\includegraphics[width=0.55\linewidth]{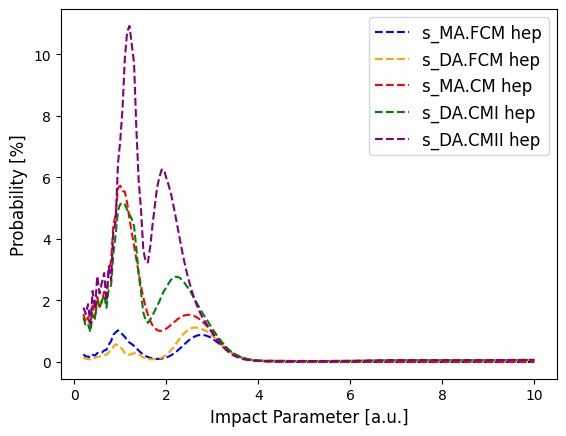}
	\caption{He$^{+}$-Ne$_{2}$ ion-dimer collisions in parallel orientation at 10 keV/amu; Single $2s$ electron removal probability comparison for: MA.FCM (blue), DA.FCM (yellow), MA.CM (red), DA.CMI (green) and DA.CMII (violet).}
	\label{fig4}
\end{figure}

As can be seen in figure~\ref{fig4}, when compared to figures~\ref{fig1} and~\ref{fig2} there is a far lower probability to remove a $2s$ electron with a He$^{+}$ projectile than with an alpha particle. The FCM probabilities are particularly small, but this model can be criticized for being unrealistic for a singly-charged projectile given that it allows for unphysical two-electron capture. Similar to what is seen in figure~\ref{fig2}, the DA.CMII probability is the largest one. 

Still, the total DA.CMII cross-section for $2s$ electron removal is much lower than for the case of He$^{2+}$ impact, (0.37\textup{~\AA}$^{2}$ vs 1.47\textup{~\AA}$^{2}$). On the other hand, a perfect ICD yield would be associated with it. This means that although the chance of obtaining Ne$^{+}$+Ne$^{+}$ fragmentation is lower, when it does occur it will always be associated with ICD. This is an extremely interesting feature of using this particular projectile to generate ICD in ion-dimer collisions. Experimental results obtained by Kim \textit{et al}~\cite{Kim13} using a He$^{+}$ projectile with a neon dimer system at 125 and 162.5 keV/amu show evidence of a strong ICD yield. However, given that direct ionization channels are open at those higher energies, direct two-electron removal processes contribute to Ne$^{+}$+Ne$^{+}$ fragmentation in the scenarios studied there. Experimental investigations of ICD generation using a He$^{+}$ projectile at lower energies, where capture strongly dominates, will provide interesting insights into the results presented in this paper.

\section{Concluding remarks}
\label{sec:conclusions}
We have studied the one- and two- electron removal processes associated with ICD, CE, and RCT in the Ne$^+$ + Ne$^+$ fragmentation channel in He$^{2+}$ and He$^{+}$-Ne$_2$ collisions for projectiles traveling parallel to the dimer axis at a collision energy of 10 keV/amu. We expanded upon the calculations presented in~\cite{Kirchner_2021} by implementing a determinantal final-state analysis that takes the Pauli principle into account and by considering the effects of the change in projectile charge due to electron capture. We determined that the various analyses and approaches demonstrate that overall similar results are obtained regardless of the method, and that the Ne($2s$) electron removal is an excellent ICD generating process. We also found that a He$^{+}$ projectile removing a Ne($2s$) electron when considered in the capture model, results in a pure ICD yield in the Ne$^+$ + Ne$^+$ fragmentation channel, albeit with a lower total cross-section than an alpha particle projectile. These results warrant further experimental investigations.

Future theoretical work will be concerned with carrying the models and analyses presented here over to argon dimer targets. While ICD has been observed in Ar$_{2}$, the situation is different from Ne$_{2}$ in that the simplest one-electron inner-valence electron removal process does not provide enough energy for facilitating ICD~\cite{Kim13}. We will also look into generalizing the capture model to make it applicable to other projectile-dimer orientations and the calculation of orientation averaged total cross-sections. We reiterate, however, that the effects of the model are most clear-cut in the parallel orientation studied here and experimental efforts to validate (or refute) it should focus on this scenario.

\newpage
\begin{acknowledgments}
Financial support from the Natural Sciences and Engineering Research Council of Canada (NSERC) 
(RGPIN-2019-06305) is gratefully acknowledged. 
\end{acknowledgments}

%

\bibliography{icd}

\end{document}